\documentclass[aps,amsfonts,nofootinbib,preprintnumbers,nobalancelastpage]{revtex4}

\usepackage[english]{babel}
\usepackage{amsmath,amssymb}
\usepackage[dvips]{graphicx}
\usepackage[sort&compress]{natbib}
\usepackage{bbm}
\usepackage{color}
\usepackage{ulem}
\usepackage{footnote}

\begin{document}


\title{Signals of Two Universal Extra Dimensions at the LHC }

\author{G. Burdman}
\email{burdman@if.usp.br}
\affiliation{Instituto de F\'{\i}sica,
             Universidade de S\~ao Paulo, S\~ao Paulo -- SP, Brazil.}

\author{O.\ J.\ P.\ \'Eboli}
\email{eboli@fma.if.usp.br}
\affiliation{Instituto de F\'{\i}sica,
             Universidade de S\~ao Paulo, S\~ao Paulo -- SP, Brazil.}

\author{D. Spehler}
\email{spehler@iphc.cnrs.fr}
\affiliation{%
Universit\'e de Strasbourg, IPHC, 23 rue du Loess 67037 Strasbourg, France
}

\begin{abstract}

  Extensions of the standard model with universal extra dimensions are
  interesting both as phenomenological templates as well as
  model-building fertile ground. For instance, they are one the
  prototypes for theories exhibiting compressed spectra, leading to
  difficult searches at the LHC since the decay products of new states
  are soft and immersed in a large standard model background.  Here we
  study the phenomenology at the LHC of theories with two universal
  extra dimensions.  We obtain the current bound 
  by using the production of 
  second level excitations of electroweak gauge bosons decaying to a
  pair of leptons and study the reach of the LHC Run~II in this
  channel. We also introduce a new channel originating in higher
  dimensional operators and resulting in the single production of a
  second level quark excitation. Its subsequent decay into a hard jet
  and lepton pair resonance would allow the identification of a more
  model-specific process, unlike the more generic 
  vector resonance signal.  We show that the sensitivity of this
  channel to the compactification scale is very similar to the one
  obtained using the vector resonance.

\end{abstract}


\maketitle

\renewcommand{\baselinestretch}{1.15}
\section{Introduction}

After the discovery of the Higgs boson~\cite{atlas,cms}, the CERN
Large Hadron Collider (LHC) is probing a new energy window, enlarging
its potential to search for new physics.  Although the Higgs boson
completes the standard model (SM) into a renormalizable, spontaneously
broken gauge theory in agreement with all experimental
data~\cite{testSM,HiggsProp}, there are many questions that remained
unanswered. Chief among them is the hierarchy problem, which would
require new physics at the TeV scale.  Moreover, up to now no
definitive signal of new physics has been observed, suggesting that
either the new physics is heavy or it is hidden by some mechanism,
such as the existence of a compressed spectrum or SM partners without
SM quantum numbers. \smallskip

Extensions of the SM that address the hierarchy problem typically
explain the Higgs mass by one of two mechanisms: either supersymmetry
is present not far from the weak scale or the Higgs is protected by a
spontaneously broken global symmetry under which it is a
pseudo-Nambu-Goldstone boson.  In both cases it is becoming
increasingly necessary to explain the absence of signals at hadron
colliders, particularly Run I at the LHC. For instance, in
supersymmetric theories it is possible to imagine that the spectrum of
new particles is dominated by states only coupled to the third
generation~\cite{naturalsusy} or that it is too compressed to result
in large enough transverse momenta or missing
$E_T$~\cite{compressedsusy}. It is also possible to build models where
the partners of the top quark are not charged under the SM
color~\cite{foldedsusy,twinhiggs}, making their observation more
difficult at hadron colliders. \smallskip

An alternative way to introduce new physics at the TeV scale is to
assume that the SM propagates in extra compact
dimensions~\cite{ued}. When supplemented with orbifold boundary
conditions, it is possible to obtain the SM as the zero-mode
spectrum. At a minimum, the new physics comes in the Kaluza-Klein
tower of excitations. Although generically, extra dimensional theories
do not solve the (little) hierarchy problem by just lowering the
cut-off, it is possible to think that they represent a new strongly
coupled sector at the TeV scale. Theories with one extra dimension
compactified on an orbifold ($S_1/Z_2$) have been thoroughly
studied~\cite{Cheng:2002ab,Edelhauser:2013lia,
  Choudhury:2016tff,Cacciapaglia:2013wha,Belyaev:2012ai,Servant:2014lqa}. 
When KK parity is assumed to be respected by the boundaries,
the resulting spectrum includes a dark matter candidate~\cite{Servant:2002aq}. Their
collider phenomenology is then similar to that of supersymmetric
extensions of the SM~\cite{Cheng:2002ab}, with cascades typically
resulting in large missing transverse momentum.  The second KK
excitation in these five-dimensional (5D) theories has a mass very
close to twice the mass of the first excitation. Thus, production of
this level-2 excitation will likely result in decays to two level-1
states, leading to a signal which is difficult to identify at hadron
colliders.  \smallskip

Theories with two universal extra dimensions can be similarly
constructed~\cite{Burdman:2005sr,Ponton:2005kx}. On the other hand,
there are some important differences. From the outset, we notice that
level-2 KK excitations have masses that are $\sqrt{2}$ times the
level-1 ones. Then, level-2 states cannot decay via the tree level
couplings that preserve KK number, but decay through one-loop
generated, KK parity preserving couplings~\cite{Burdman:2006gy} to SM
particles, 
leading to more identifiable signals at the LHC.  \smallskip

In the present paper we consider the six-dimensional standard model
(6DSM)~\cite{Burdman:2006gy} as an example of a model that possesses a
compressed spectrum.  KK number conservation implies that level-1 KK
modes can only be produced in pairs. Their decays contains soft
leptons, jets and missing energy making their discovery very
difficult~\cite{Cheng:2002ab,Choudhury:2011jk}. On the other hand, level-2 KK states
can be singly produced in the $s$-channel~\cite{Burdman:2006gy}
and can decay into pairs of SM particles due to KK-number violating
interactions.  \smallskip

We will examine the LHC potential for studying the 6DSM through the
inclusive search for new narrow vector resonances decaying into lepton
pairs ($e^\pm$ or $\mu^\pm$) that takes place via its s-channel
production
\begin{equation}
    p p \to W^{3(1,1)}_\mu/B_\mu^{(1,1)} + X \to \ell^+ \ell^- + X \;,
\label{eq:leptons}
\end{equation}
as well as through the resonant production of $(1,1)$ KK  quarks
in the channel
\begin{equation}
   p p \to Q^{(1,1)} \to \ell^+ \ell^- + \hbox{jet} \; ,
\label{eq:kkq}
\end{equation}
where $Q^{(1,1)}$ stands for the $(1,1)$ KK quarks and
$W^{3(1,1)}_\mu$ and $B_\mu^{(1,1)}$ are the level-2 vector states of
the SM electroweak gauge bosons. \smallskip

We will show that the level-2 excitations of the electroweak gauge
bosons in Eq.~(\ref{eq:leptons}) provide the best bound for the
compactification scale $R$ from available Run I and II data.  We will
then study the reach of the LHC in Run~II. For this purpose, we first
use the s-channel resonance going into lepton pairs of
Eq.~(\ref{eq:leptons}), but we also add a previously unexplored
channel: the single production of a $(1,1)$ quark as in
Eq.~(\ref{eq:kkq}).  Although we will see that the LHC reach in $1/R$
is similar in this second channel as it is in the dilepton case, this
addition constitutes a more model-specific signal since it comes from
KK-number violating higher dimensional operators typically present in
extra-dimensional theories, whereas the channel in
Eq.~(\ref{eq:leptons}) is omnipresent in SM extensions. The level-2
quark decays into one of the dilepton resonances and a hard jet,
allowing for its reconstruction.  Both signals combined would provide
an interesting pattern pointing the direction to further searches and
model building based on this simple 6DSM construction. \smallskip

The rest of the paper is organized as follows: in Section~\ref{dim6}
we review the 6DSM, focusing on the main phenomenological facts such
as the spectrum and couplings, resulting in specific decay
patterns. In Section~\ref{bounds} we obtain the current bound on $1/R$
using the available LHC Run I and II data. We study the potential of
the LHC Run~II data in these channels in Section~\ref{lhcpot} and
conclude in Section~\ref{conclusions}.

\section{Six-dimensional standard model}
\label{dim6}

We consider the six-dimensional standard model as defined
in Ref.~\cite{Burdman:2006gy} where the two extra dimensions form a
square $ 0 \le x^4,x^5 \le \pi R$, and  are compactified by identifying
pairs of adjacent sides, the so called ``chiral
square''~\cite{Burdman:2005sr,Ponton:2005kx}.  The extra dimensional space is
symmetric under reflections with respect of the center of the square,
being this the KK parity symmetry that we label $Z^{KK}_2$.
\smallskip

In order to ensure 6D anomaly cancelation~\cite{Dobrescu:2001ae}, the
weak-doublet quarks have the opposite 6D chiralities than the singlet
ones, {\it i.e} labeling the 6D chiralities as $\pm$ we have
$Q_+ = (U_+, D_+)$, $U_-$ and $D_-$. In addition to these states the
model contains 6D gauge bosons and leptons. The Kaluza-Klein (KK)
expansion of a six-dimensional field $\Phi$ possessing a zero
mode can be written as
\begin{equation}
 \Phi = \sum_{j,k} \left ( \cos \frac{j x^4 + k x^5}{R} + \cos \frac{k
     x^4 - j x^5}{R} \right ) \frac{\Phi^{(j,k)}(x^\mu)}{ \pi R (1 +
   \delta_{j,0})} 
\; ,
\end{equation}
where the KK numbers $j$ and $k$ are integers satisfying $j \ge 1$ and
$k\ge 0$ or $j=k=0$. At tree level, the masses of the 4D KK modes
$\Phi^{(j,k)}$ are
\begin{equation}
  M_{j,k} = \sqrt{j^2 + k^2}~\frac{1}{R} \; .
\end{equation}
In the 6DSM each 6D gauge boson ($V$) decomposes into a tower of 4D
spin-0 ($V_H^{(j,k)}$) and spin-1 ($V_\mu^{(j,k)}$) fields~\cite{Burdman:2005sr,Ponton:2005kx}. On the
other hand, the 6D lepton and quark fields give rise to a tower of
massive vector-like 4D fermions and a chiral zero mode that we
identify with the known fermions.  \smallskip

In addition to the bulk action, the theory admits operators localized
at the fixed points of the chiral square, {\it i.e.}  the points
  $(0,0)$, $(\pi R, \pi R)$ and $(0,\pi R) \equiv(\pi R,0)$.  These
are induced by loops involving the 6D bulk
interactions~\cite{Georgi:2000ks}, as well as by physics above the
cutoff scale. Generically, we have~\cite{Burdman:2006gy}
\begin{equation}
\int_0^{\pi R} dx_4\int_0^{\pi R} dx_5 \left\{ {\cal L}_{\rm bulk} +
    \delta(x_4)\delta(x_5-\pi R)\,{\cal L}_2
+ \left[\delta(x_4)\delta(x_5)+\delta(x_4-\pi R)\delta(x_5-\pi
   R)\right]\,{\cal L}_1 \right\}~,
\label{lag4d} 
\end{equation}
which reflects the identification of the points at $(0,\pi R)$ and
$(\pi R,0)$, as well as the KK parity, $Z_2^{KK}$ resulting in
identical operators at $(0,0)$ and $(\pi R,\pi R)$.  Here
${\cal L}_{\rm bulk}$ is the bulk 6D Lagrangian including all the SM
fields kinetic terms as well as the appropriate Yukawa couplings and
Higgs potential needed in order to obtain the SM as the zero-mode
spectrum.  The terms ${\cal L}_1$ and ${\cal L}_2$ contain all
possible localized operators consistent with the 4D symmetries of the
SM as well as the pieces that correspond to the motion along
the two extra dimensions.  For instance, the lowest-dimension
localized operator involving the 6D $U_-$ quark field is
\begin{equation}
\frac{C_{pU}}{\Lambda^2}\,\left(\bar{U}_{-R}\,i\Gamma_\mu D^\mu U_{-R}
  + \bar{U}_{-L}\,i\Gamma_\mu D^\mu U_{-L} \right) +
\left(\frac{C'_{pU}}{\Lambda^2}\,\bar{U}_{-R}\,i\Gamma_\ell U_{-L} +
  {\rm h.c.}\right) ~,
\label{ulockinterms}
\end{equation}
where $\Gamma_\mu$ with $\mu=0,1,2,3$, and $\Gamma_\ell$ with
$\ell=4,5$ are $8\times 8$ anti-commuting matrices defining the
Clifford algebra in 6D, $D_\mu$ and $D_\ell$ are covariant
derivatives, and the order one coefficients $C_{pU}$ and $C'_{pU}$ are
partly determined by the physics above the cutoff $\Lambda$, plus
renormalizations arising below it. The index $p=1,2$ refers to
operators belonging to ${\cal L}_1$ and ${\cal L}_2$.  Similarly,
localized operators containing the 6D gluon field give rise to the
following lowest mass dimension operators
\begin{equation}
-\frac{1}{4}\,\frac{C_{pG}}{\Lambda^2}\,G_{\mu\nu}G^{\mu\nu}
-\frac{1}{2} \frac{C'_{pG}}{\Lambda^2} \,(G_{45})^2\;.
\label{glockinterms}
\end{equation}

\begin{figure}[htb!]
  \centering
  \includegraphics[width=0.45\textwidth]{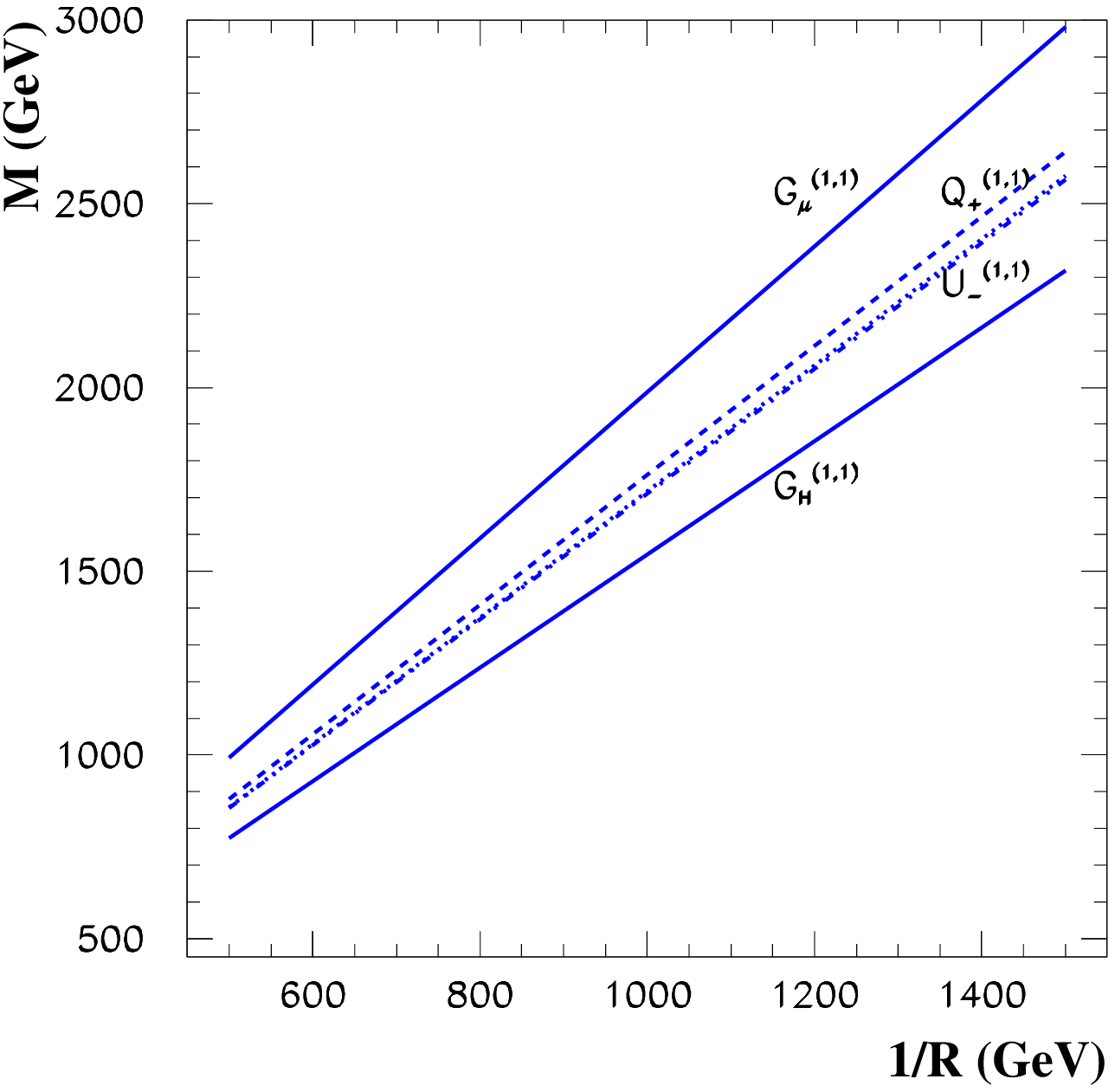}
  \includegraphics[width=0.45\textwidth]{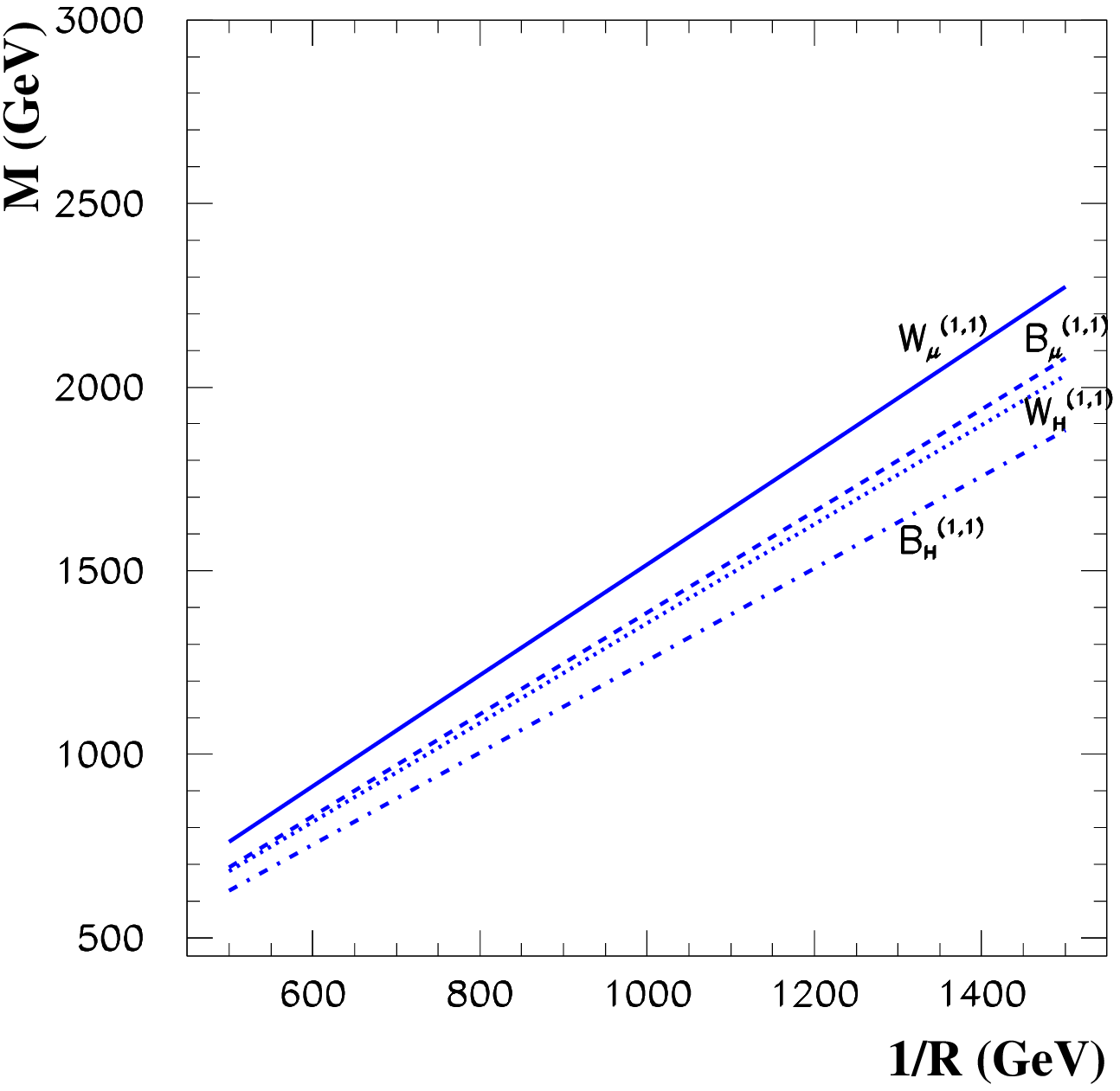}
  \caption{ Mass spectrum of the $(1,1)$ KK states as a function of
    $1/R$. In the left panel we depict the masses of the strongly
    interacting states while the right panel contains the $(1,1)$ KK
    states associated to the 6D electroweak gauge bosons.  Notice
      that the states $U_-^{(1,1)}$ and $D_-^{(1,1)}$ are almost
      degenerate.  }
\label{fig:1}
\end{figure}

The presence of these localized operators results in interactions
violating KK number conservation but respecting the $Z_2^{KK}$
parity. Additionally, these operators result in corrections to the
masses of the KK tower that would depend on the gauge charges of the
fields.  For illustration, we present in Figure~\ref{fig:1} the
expected mass spectrum of the $(1,1)$ KK states where we used the
expressions given in Ref.~\cite{Burdman:2006gy,Cheng:2002iz} and we
chose the cut-off scale $\Lambda = 10/R$. The splitting in the
spectrum does allow for KK-number conserving cascade decays of
strongly produced KK states into lighter ones, albeit with difficult
to observe final states. \smallskip

More promising signals result from the KK-number violating
interactions induced by localized operators such as the ones in
Eqs.~(\ref{ulockinterms}) and (\ref{glockinterms}).  The $Z^{KK}_2$
implies that the sum over all $j$ and $k$ numbers must be even in
interactions among the KK states.  Bulk interactions do not generate
interactions between a KK mode and two zero modes since they respect
KK number. However, localized operators can give rise to them
 allowing the decay of a KK state directly into two SM
particles~\cite{Burdman:2006gy}. For instance, KK-number violating
couplings between a massive KK gluon and SM quarks are described by
the operator~\cite{Burdman:2006gy}
\begin{equation}
g_s\,C_{j,k}^{qG}   \left( \bar{q} \gamma^\nu T^a q \right )
G^{(j,k)a}_\nu \;,
\label{z2violren}
\end{equation}
where $T^a$ are the $SU(3)_c$ generators in the fundamental
representation.  The coefficient $C_{j,k}^{qG}$ receives contributions
from the localized operators in Eqs.~(\ref{ulockinterms}) and
(\ref{glockinterms}) through the renormalizations of the quark and
gluon lines they induce.  There are similar interactions for the
electroweak states $W_\mu^{(j,k)a}$ and $B_\mu^{(j,k)}$ with the
natural adjustments for a different gauge group. \smallskip

As shown in Ref.~\cite{Burdman:2006gy} the existence of these
KK-number violating interactions such as in Eq.~(\ref{z2violren}) has
important consequences for the search of $(1,1)$ KK states of the
gauge bosons.  Since their masses are $\sqrt{2}$ times the $(1,0)$
masses, they cannot decay into them. Then, although the couplings in
Eq.~(\ref{z2violren}) are volume suppressed when compared with the
KK-conserving ones, they determine the $(1,1)$ decay channels.
\smallskip

In addition, here we show that there are operators of higher mass
dimensions that are potentially as important as these in the
phenomenology at the LHC.  In particular we study localized operators
that allow the direct coupling of $(1,1)$ KK quarks to pairs of SM
particles. Of interest to us here is the one-loop induced process
\begin{equation}
      qg \to   Q^{(1,1)}  \;,
\label{Qqg}
\end{equation}
where $Q$ stands for any of the $(1,1)$ KK states of the quarks while
$q$ ($g$) represents a SM light quark (gluon). \smallskip

In order to generate a process like Eq.~(\ref{Qqg}) via localized
operators we would need to go to operators of higher mass dimensions
than the ones leading to Eq.~({\ref{z2violren}).  The reason is that
  to obtain a non-diagonal (in KK number) gluon couplings to quark
  fields, these cannot come from kinetic-like localized operators
  since these interactions will always be diagonal due to gauge
  invariance. On the other hand, higher dimensional localized
  operators such as
\begin{equation}
{\cal O}_1=\overline{U}\,\Gamma^\mu \, D^\nu U\, G_{\mu\nu}\quad 
\hbox{ and } \quad
{\cal O}_2=\overline{U}\,\sigma^{\mu\nu} U\, G_{\mu\nu} 
\label{hdops}
\end{equation} 
will lead to processes like Eq.~(\ref{Qqg}).  \smallskip

Expanding the 6D fields in Eq.~(\ref{hdops}) into KK modes results in
the effective Lagrangian for the process of interest given by
\begin{equation}
  f_1 \overline{Q^{(1,1)}} \gamma^\mu D^\nu T^a q~  G^{a}_{\mu\nu}
+   f_2 \overline{Q^{(1,1)}} \sigma^{\mu\nu} T^a q~  G^{a}_{\mu\nu}
\label{eq:Leff}
\end{equation}
where $f_{1,2}$ are functions of the momenta of the particles and $R$.
Notice that the effective operators in Eq.~(\ref{eq:Leff}) allow the
single production of $(1,1)$ KK quarks that has the potential of
enlarging the LHC capabilities to search for these particles as we
show below.\smallskip

\begin{figure}[htb!]
  \centering
  \includegraphics[width=0.65\textwidth]{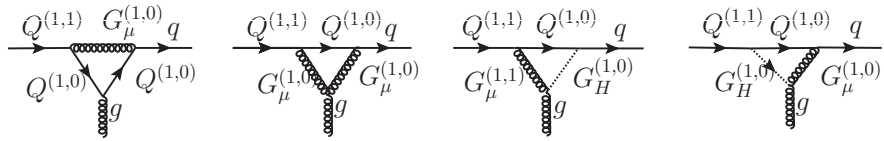}
  \caption{Feynman diagrams contributing to the one-loop process
    $Q^{(1,1)} \to q g$.  }
\label{fig:feyn}
\end{figure}

We can estimate the coefficients of the operators in Eq.~(\ref{hdops}) by computing the
one-loop contributions to them coming from KK excitations through bulk
vertices that respect KK number.  In Figure~\ref{fig:feyn} we show the
Feynman diagrams corresponding to these one-loop contributions, which
are finite.  The Wilson coefficients of the operators given by
  Eq.~(\ref{eq:Leff}) are given by
\begin{equation}
\begin{array}{l}
f_1 =\frac{\alpha_s^{3/2}}{\sqrt{4\pi}}\,\left\{ (C_2(R)-C_2(G)) \left(8C_{23}-4C_{22}-4C_{21}\right)
        +2C_2(R)C_0 +
        C_2(G)\left(-C_0+C_{12}-C_{11}\right)\right\} \;,
\\
\\
f_2=
       \frac{\alpha_s^{3/2}}{\sqrt{4\pi}}\,\left\{4\left(C_2(R)-C_2(G)\right)\left(C_{23}-C_{21}\right)+2C_2(R)C_0
       +\left(-3C_0-3C_{12}+C_{11}\right)\right\} \;,
\end{array}
\label{f1f2}
\end{equation}
where the $C_X$ are the Passarino-Veltman
functions~\cite{Passarino:1978jh} evaluated at
$C_x(0,0,M_{11}^2, M_{10}^2,M_{10}^2)$, with the $M_{ij}$ the masses
of the $(i,j)$ KK state, and $C_2(R)=4/3$ and $C_2(G)=3$ are the
Casimir invariants of the fundamental and adjoint representations of
$SU(3)$ respectively.  Although the coefficient functions $f_1$ and
$f_2$ generally receive additional contributions from the UV, we will
estimate their size by the one loop contributions from
Figure~\ref{fig:feyn} and detailed in Eq.~(\ref{f1f2}). \smallskip

In what follows we study the phenomenology of the $(1,1)$ KK modes.
Although the bulk KK-number conserving interactions mediate the decay
of a given $(1,1)$ state into a lighter $(1,1)$ state plus a SM
particle, localized KK-number violating interactions open up decays of
$(1,1)$ states to pair of SM particles~\cite{Burdman:2006gy}. These
are often the preferred modes, as it can be seen in
Table~\ref{tab:bw}, where the $B_\mu^{(1,1)}$ and $W^{3(1,1)}_\mu$
branching ratios that are clearly dominated by KK-number violating
interactions.  Moreover, the KK-number violating interactions are also
responsible for the decays of the spin-0 adjoint states $G_H^{(1,1)}$, $W_H^{3(1,1)}$ and $B_H^{(1,1)}$
into top-quark pairs since their couplings to  fermions are
proportional to the fermion mass~\cite{Burdman:2005sr}.
\smallskip

\begin{table}
\begin{tabular}{|c||c|c|}
\hline
decay mode & $B_\mu^{(1,1)}$ & $W^{3(1,1)}_\mu$ 
\\
\hline
$t \bar{t}$ & 29\% & 15\%
\\
\hline
$ b \bar{b}$ & 7\% & 16\%
\\
\hline
light dijet  & 60\% & 50\%
\\
\hline
$\sum \ell^+ \ell^-$ & 3\% & 0.05\%
\\
\hline
$\sum \nu \bar{\nu}$ & 1\% & 0.05\%
\\
\hline
$\sum L^{(1,1)} + \ell$ & --- & 19\%
\\
\hline
\end{tabular}
\caption{ Two-body decays of $B_\mu^{(1,1)}$ and $W^{3(1,1)}_\mu$ and
  respective branching ratios for   $1/R= 1 $ TeV. }
\label{tab:bw}
\end{table}

On the other hand the decay of the $(1,1)$ KK quarks takes place
mostly through KK-conserving interactions as can be seen in
Table~\ref{tab:quarks}. In fact, we verified that the branching ratio
of the $(1,1)$ KK quarks into quark-gluon pairs via the interactions
in Eq.~(\ref{eq:Leff}) is negligible. Thus, we can make use of the
single production of the $(1,1)$ quark through Eq.~(\ref{Qqg})
followed by its decays into $W^{(1,1)}_\mu$ plus jet and
$B^{(1,1)}_\mu$ plus jet. This provides an additional search
channel for a 6DSM signal: s-channel resonant production of
$(1,1)$ KK quarks.
\smallskip

\begin{table}
\begin{tabular}{|c||c|c|c|}
\hline
decay mode & $Q_+^{(1,1)}$ & $U_-^{(1,1)}$ & $D_-^{(1,1)}$
\\
\hline
$G_H^{(1,1)} q $& 42\% & 64\% & 87\%
\\
\hline
$W_H^{3(1,1)} q$ & 24\% & -- & --
\\
\hline
$\sum_j W_\mu^{j(1,1)}q$ & 32\% & -- & --
\\
\hline
$B_H^{(1,1)}q$ & 0.4\% & 12\% & 4\%
\\
\hline 
$B_\mu^{(1,1)}q$ & 0.8\% & 24\% & 9\% 
\\
\hline
\end{tabular}
\caption{Branching ratios of $(1,1)$ KK quarks of the first two
  generations for $1/R = 1$ TeV.}
\label{tab:quarks}
\end{table}


 Taking into account that the branching ratios shown in
  Tables~\ref{tab:bw} and \ref{tab:quarks} for $1/R= 1$ TeV do not
  change much when we vary $R$, we see that inclusive channels
  containing leptons  will lead to the first signal
  of the existence of the 6DSM.
 In the next section we will obtain
  bounds on $1/R$ using the lepton channels with the available 
    LHC Run~I and II data at $\sqrt{s}=8$ and $13$ TeV respectively.
    We will make then sensitivity predictions for the LHC at
    $\sqrt{s}=13~$TeV in this channel for larger integrated
    luminosities. But in addition, we will also make use of the
  singly-produced $(1,1)$ KK quark with its subsequent KK-number
  conserving decays to a charged lepton pair and a hard jet to show
  that this is a complementary channel which can prove important in
  pinning down the origin of the signals being observed beyond the rather
  ubiquitous 
  vector resonance signal. \smallskip


\section{Present bounds on the 6DSM} 
\label{bounds}

We start by making use of the LHC Runs~I and II available data in
order to extract the current bound on $1/R$ in the 6DSM.  In
particular, the CMS collaboration searched for narrow resonances ($V$)
decaying into $e^+e^-$ or $\mu^+\mu^-$ pairs at the center-of-mass
energies of 8 and 13 TeV and an integrated luminosities of 20.6
  and 2.6 fb$^{-1}$ respectively~\cite{CMS:dileptons,CMS:2015nhc}.
The analysis was based in the ratio between inclusive production cross
sections
\begin{equation}
R_\sigma = \frac{\sigma(pp \to V + X \to \ell^+\ell^- +X) }
{\sigma(pp \to Z + X \to \ell^+\ell^- +X)}
\label{eq:r}
\end{equation}
where $\ell=e$ or $\mu$ and the $V$ production cross section was
obtained using a window of 40\% its mass in the 8 TeV analysis
  and the narrow width approximation in the 13 TeV one. On the other
hand, the $Z$ production cross section used a mass window of $\pm 30$
GeV in both analyses. The 6DSM contribution to $R_\sigma$
originates from the processes in Eqs.~(\ref{eq:leptons}) and
(\ref{eq:kkq}). \smallskip

In our analysis we evaluated the relevant cross sections at tree level
using the package MADGRAPH~\cite{madgraph} where the 6DSM
was inputed using FeynRules~\cite{feynrules}.
In Figure~\ref{fig:limits} we show the present CMS 8 and 13 TeV
limits on $R_\sigma$ as a function of the mass of the new narrow
resonance, as well as, the 6DSM prediction. The 6DSM cross section is
dominated by the $B_\mu^{(1,1)}$ production with a few percent
contribution from $W^{3(1,1)}_\mu$, see Table~\ref{tab:bw}. As
we can see from the left panel of this figure, the CMS Run I data lead
to the 95\% CL constraint
\begin{equation}
    M_V > 1250 \hbox{ GeV.}
\label{eq:limit:mv}
\end{equation}
that can be translated into
\begin{equation}
   \frac{1}{R} > 900 \hbox{ GeV.}
\label{eq:limit:r}
\end{equation}
On the other hand the limits originating from the 13 TeV data,
  {\it i.e.} $M_V > 1140$ GeV and $1/R > 820$ GeV are weaker than the
  ones coming from Run I,  due to the small 13 TeV integrated
  luminosity.  
Furthermore, it is interesting to notice that the above limit on $1/R$
is very close to the indirect bound that can be obtained from the
precision electroweak measurements that lead to $1/R > 920$ GeV at
95\% CL ~\cite{Kakuda:2013kba}. \smallskip

Other potentially interesting bounds can come from the spinless
adjoints $G_H^{(1,1)}$, $W_H^{3(1,1)}$ and $B_H^{(1,1)}$, which appear
in the spectrum of the 6DSM given that only one combination of
Nambu-Goldstone bosons is eaten by the KK excitations.  Since they
couple to mass they decay predominantly into top quark
pairs~\cite{Burdman:2006gy}. The ATLAS and CMS collaborations looked
for $t \bar{t}$ resonances at the LHC with a center-of-mass energy of
8 TeV~\cite{TheATLAScollaboration:2013kha,Chatrchyan:2013lca}.  We
verified that the present available limits on the production cross
section of $t \bar{t}$ resonances do not lead to competitive bounds on the 6DSM.
\smallskip

\begin{figure}[htb!]
  \centering
  \includegraphics[width=0.45\textwidth]{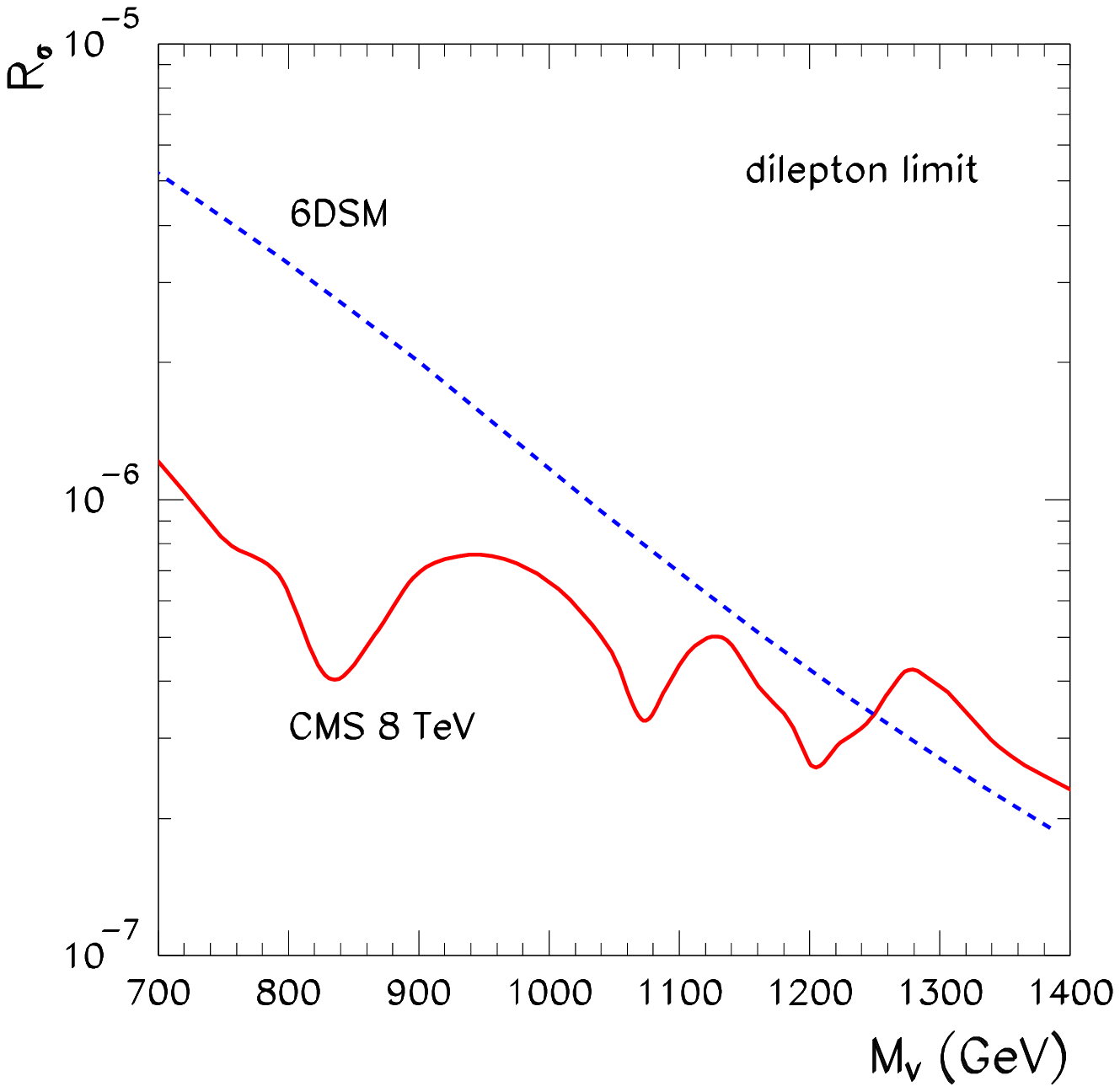}
  \includegraphics[width=0.45\textwidth]{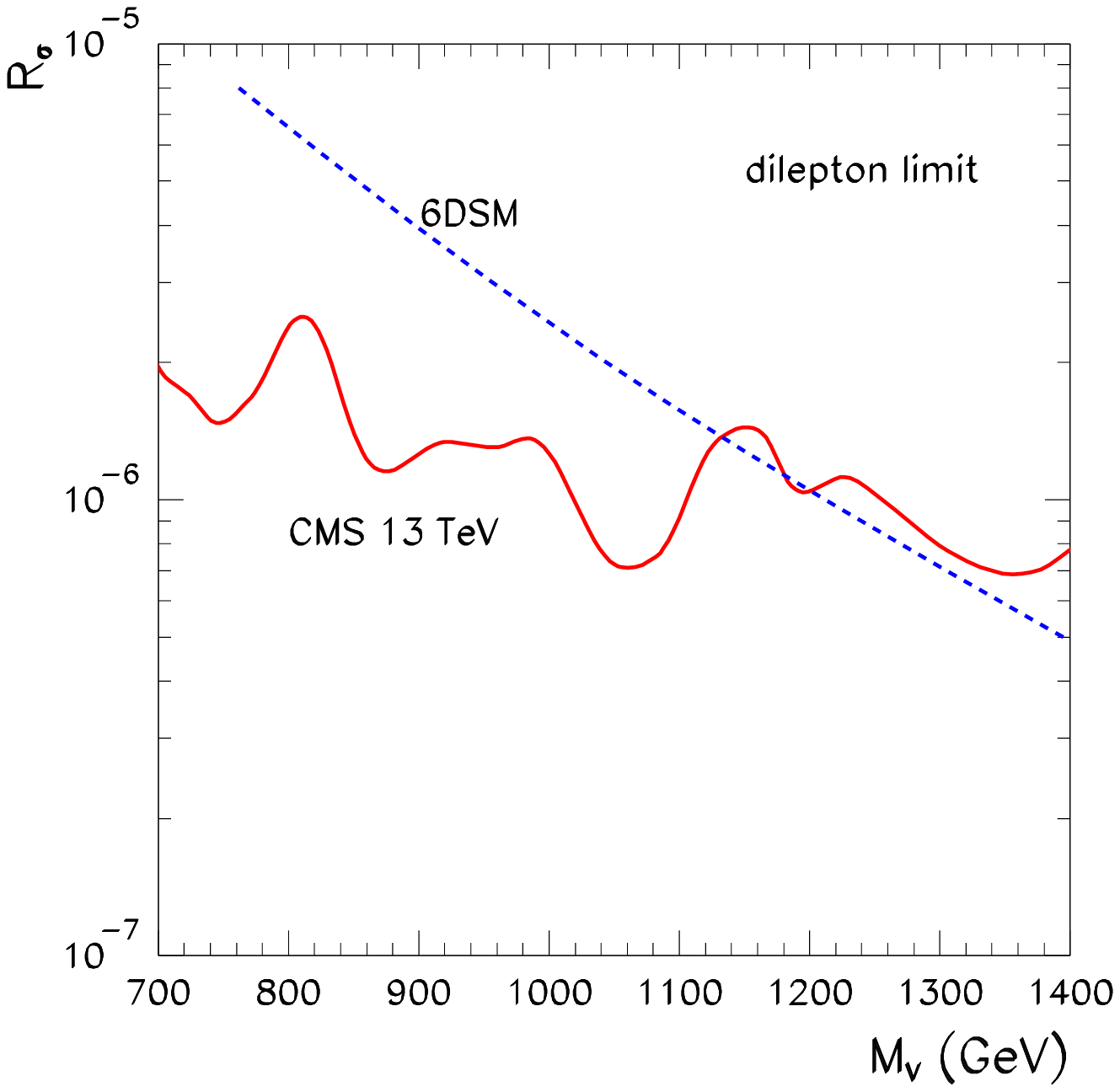}
  \caption{The solid red line stands for the CMS 95\% CL upper limit
    on the ratio $R_\sigma$ of a narrow resonance decaying $e^+ e^-$
    and $\mu^+ \mu^-$ pairs as a function of the resonance mass $M_V$
    in GeV. The dashed blue line represents the six-dimensional
    standard model prediction for $R_\sigma$. The left (right)
      panel contains the 8 (13) TeV results.} 
\label{fig:limits}
\end{figure}

\section{LHC potential to further probe the 6DSM}
\label{lhcpot}

In evaluating the LHC reach for finding the 6DSM we will consider two
strategies. The first is the search for opposite charge dilepton 
  $\ell^+\ell^-$ resonances with $\ell^\pm = e^\pm$ or $\mu^\pm$. The
  dilepton pairs originated from the s-channel production of
  $B_\mu^{(1,1)}$ and $W_\mu^{3(1,1)}$, as well as from the $(1,1)$ KK
  quarks decays into these $(1,1)$ vector resonances, followed by
  their decays to lepton pairs with the branching ratios detailed in
  Table~\ref{tab:bw}. Although these branching fractions are small,
the cleanliness of the signal allows for interesting bounds, as we saw
in the previous section for the Runs~I and II results. On the
other hand, a dilepton resonance is present in many extensions of the
SM and it would be good to have a signal that is more
model-specific. Thus, our second strategy will be to look for the
single production of a $(1,1)$ excited quark and its subsequent decay
into a jet and a lepton pair resonance, either $B_\mu^{(1,1)}$ or
$W_\mu^{3(1,1)}$. The single $(1,1)$ quark production mechanism coming
from higher dimensional operators is a sign that we would be in the
presence of non-renormalizable interactions suppressed by not such as
high scale, a typical feature of the 6DSM. In what follows we detail
these two strategies and their reach in Run~II at the LHC.

\subsection{Search for new resonances in the inclusive $\ell^+ \ell^-$ final state}

The 6DSM contributes to the inclusive production of dilepton 
  resonances through the production of the $(1,1)$ KK vector bosons
  $B^{(1,1)}_\mu$ and $W^{3~(1,1)}_\mu$ by the processes given in
  Eqs.~(\ref{eq:leptons}) and (\ref{eq:kkq}). The main SM backgrounds
  for these are the processes~\cite{CMS:2015nhc}
\begin{eqnarray}
p p & \to & \ell^+ \ell^- + X \;,
\nonumber
\\
p p & \to & W^+ W^-/Z Z \to \ell^+ \ell^- \nu_\ell \bar{\nu}_\ell \; ,
\label{eq:bckg:ll}
\\
p p & \to & t \bar{t} \to \ell^+ \ell^-  \nu_\ell \bar{\nu}_\ell + \hbox{jets}.
\nonumber
\end{eqnarray}

Initially we evaluated the LHC potential to constrain the 6DSM via the
resonance search in the dilepton channel assuming a center-of-mass
energy of 13 TeV and an integrated luminosity of 30 (100) fb$^{-1}$.
In this first scenario we assumed that the number of observed events
agrees with the SM prediction to extract the attainable 95\% CL limits
on the mass of the vector resonances or on the compactification radius
$R$. Once again we simulated the signal in Eqs.~(\ref{eq:leptons}) and
(\ref{eq:kkq}), as well as the SM backgrounds in
Eq.~(\ref{eq:bckg:ll}) at tree level using the package MADGRAPH. We
added the $\mu^+\mu^-$ and $e^+ e^- $ contributions assuming that the
muon reconstruction efficiency is close to 100\% and the
electron/positron one is 90\%. \smallskip

In our analysis of the inclusive dilepton signal we imposed very
simple acceptance cuts on the charged leptons
\begin{equation}
| \eta_\ell | < 2.5 \;\;\;\;\;\hbox{ and }\;\;\;\;\;p_T > 100 \hbox{ GeV}.
\label{eq:cuts:1}
\end{equation}
We also required that the dilepton invariant mass ($m_{\ell\ell}$)
lays in a window around the resonance mass $M_V$ whose width is 10\%
of $M_V$, {\it i.e.}
\begin{equation}
   |m_{\ell\ell} - M_V | < 0.05 \times M_V \; .
\label{eq:cuts:2}
\end{equation}

We present in Fig.~\ref{fig:lepbounds} the limit on new resonance
contributions to the inclusive cross section after cuts as a function
of $M_V$ (left panel) and reinterpreted it as a bound on $1/R$ (right panel) for integrated
luminosities of 30 (dotted blue line) and 100 fb$^{-1}$ (dashed red
line).  As we can see from this figure the LHC with a center-of-mass
energy of 13 TeV will be able to exclude at 95\% CL $(1,1)$ vector
resonances with masses up to 2.0 (2.4) TeV for 30 (100)
fb$^{-1}$. These bounds correspond to limits on $1/R$ of 1.4 (1.7) TeV
respectively. \smallskip

\begin{figure}[htb!]
  \centering
  \includegraphics[width=0.45\textwidth]{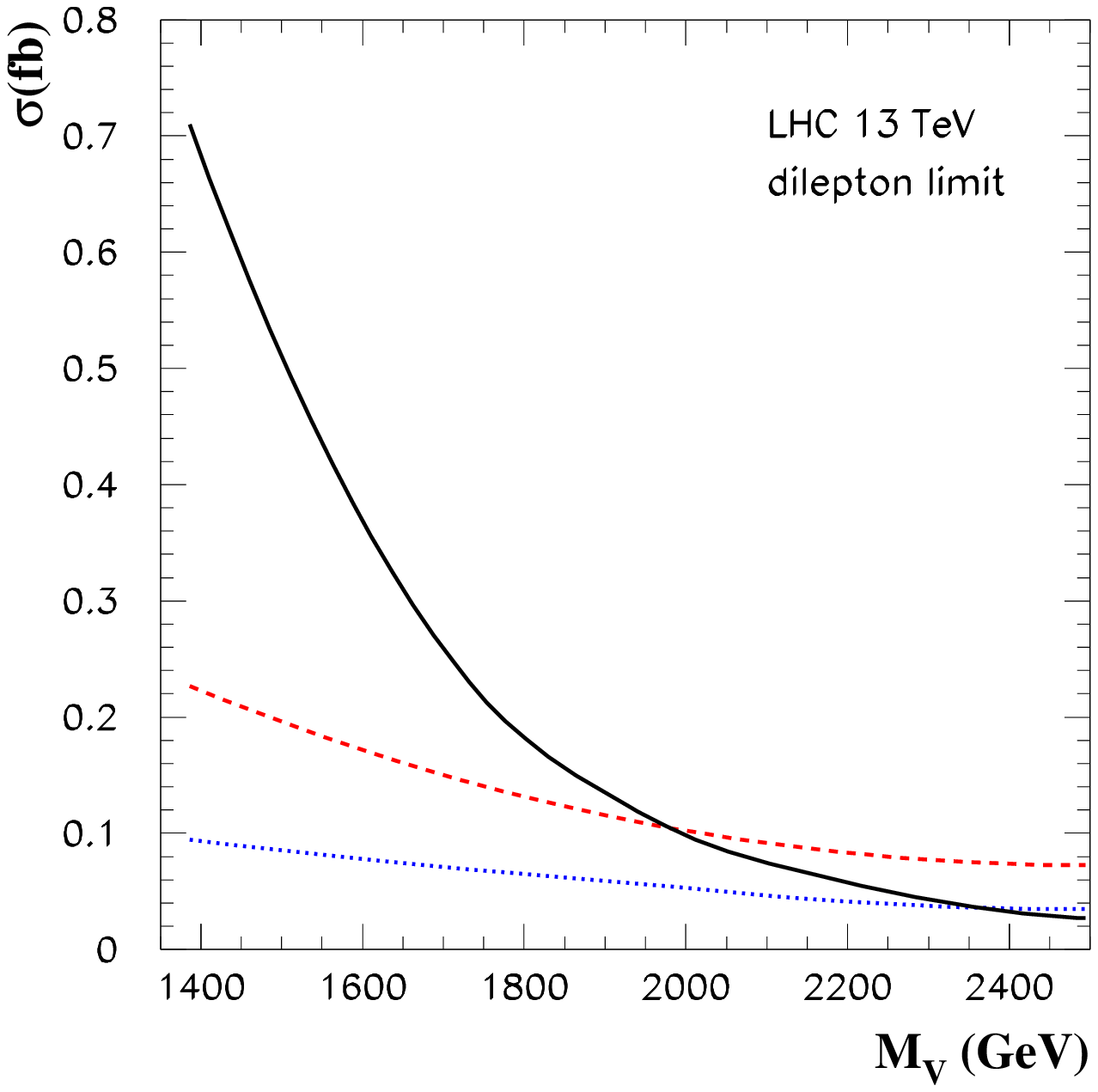}
  \includegraphics[width=0.45\textwidth]{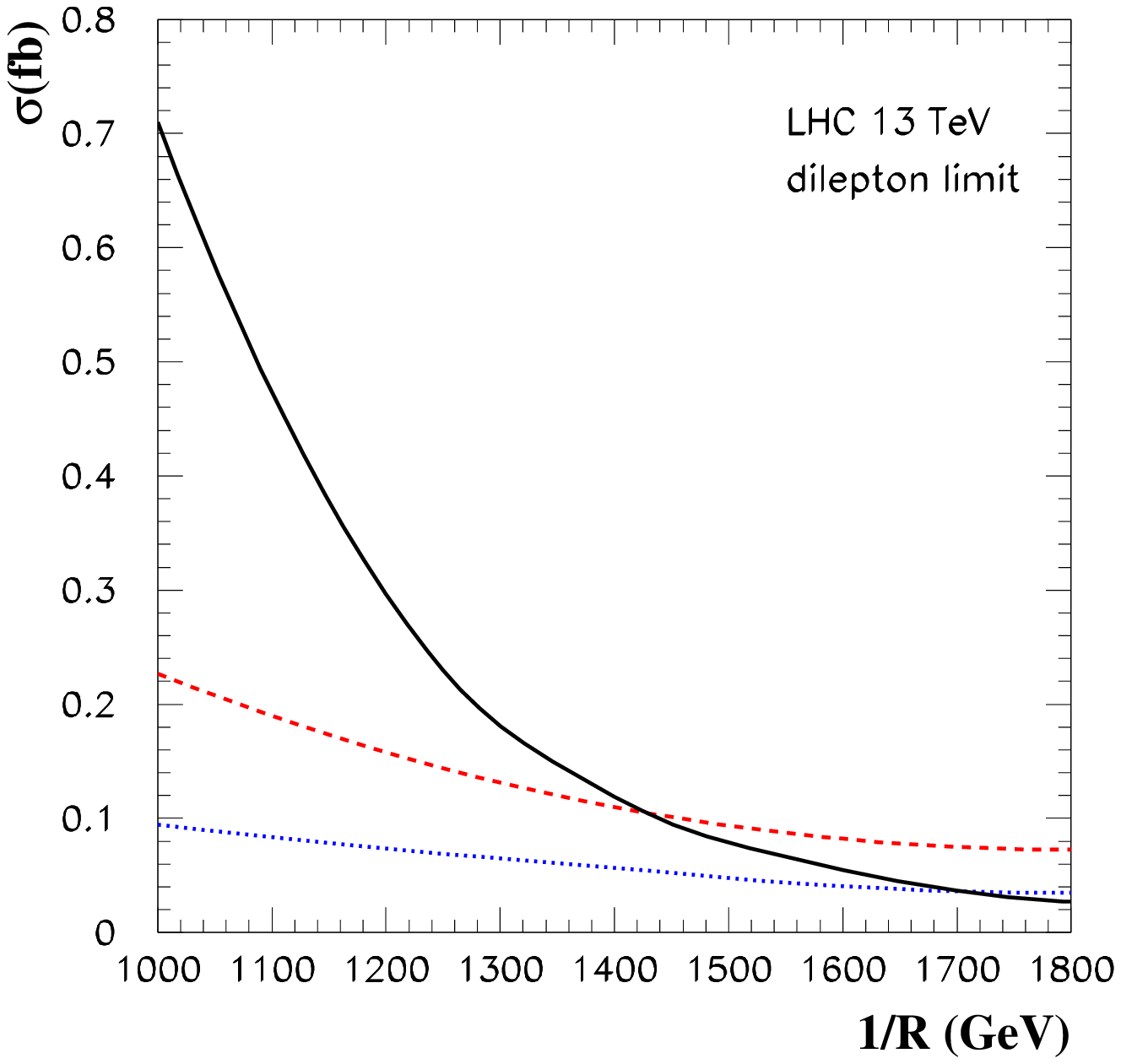}
  \caption{95\% CL attainable limits on new resonance contribution to
    the dilepton production cross section as a function of the $B^{(1,1)}_\mu$
    mass (left) or as function of $1/R$ (right).  The
     dashed red (dotted blue) curve stands for the limit on the
    production cross section after cuts assuming an integrated
    luminosity of 30 (100) fb$^{-1}$. The new contribution to the
    dilepton production cross section due to the six-dimensional
    standard model is depicted by the solid black line.  }
\label{fig:lepbounds}
\end{figure}

In order to assess the LHC discovery potential of the six-dimensional
standard model through the inclusive dilepton channel we determine the
integrated luminosity necessary for a $5\sigma$ discovery; our results
are shown in Fig.~\ref{fig:lepdisc}.  As we can see from the left
panel of this figure, the LHC can discover $B^{(1,1)}_\mu$ with masses
up to 1530 (1850) GeV for an integrated luminosity of 30 (100)
fb$^{-1}$. From the right panel of the same figure, we can see that
the LHC can unravel signals of the 6DSM for compactification scales
($1/R$) 1100 (1340) GeV for the above integrated luminosities
respectively. \smallskip

\begin{figure}[htb!]
  \centering
  \includegraphics[width=0.45\textwidth]{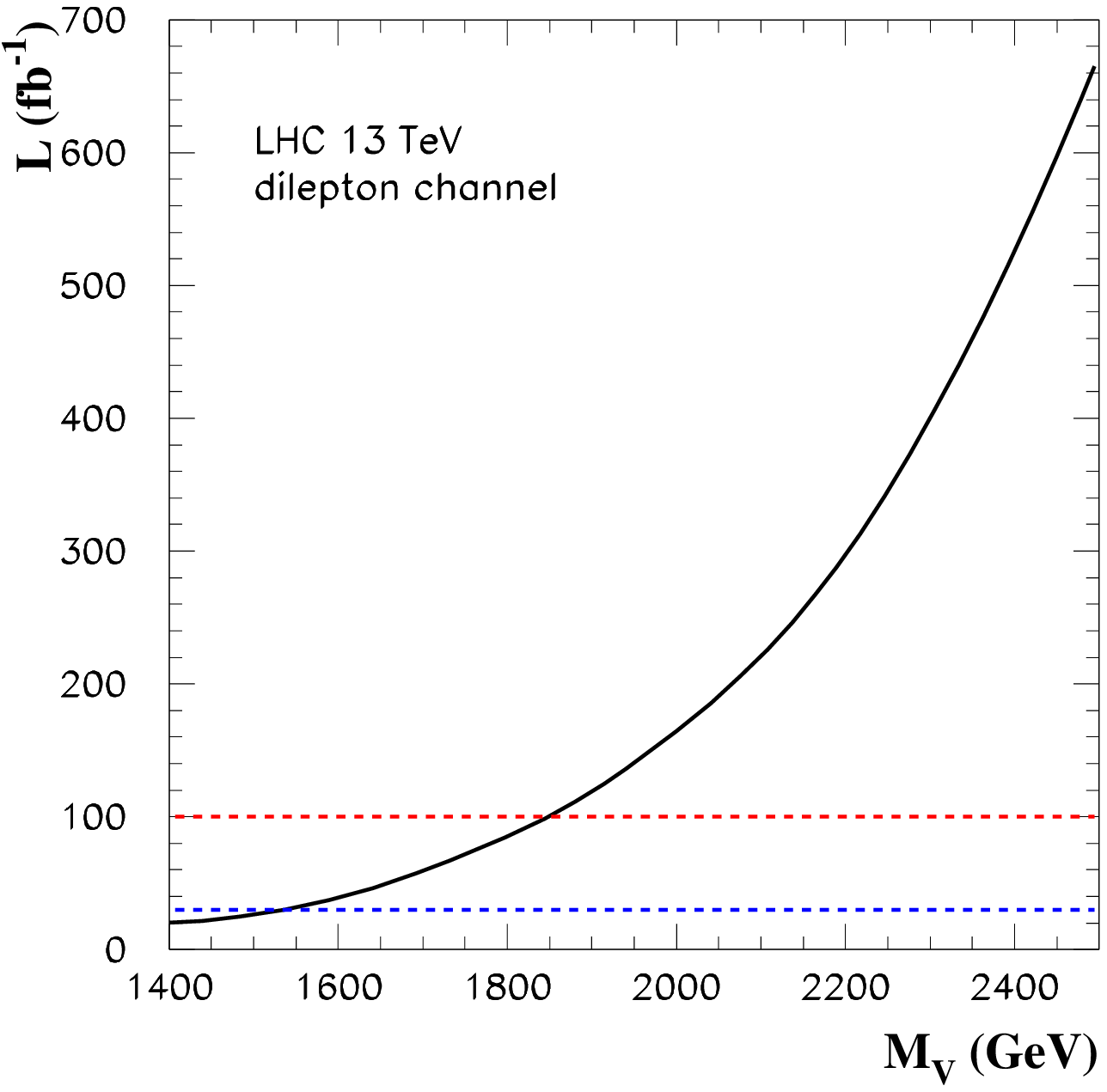}
  \includegraphics[width=0.45\textwidth]{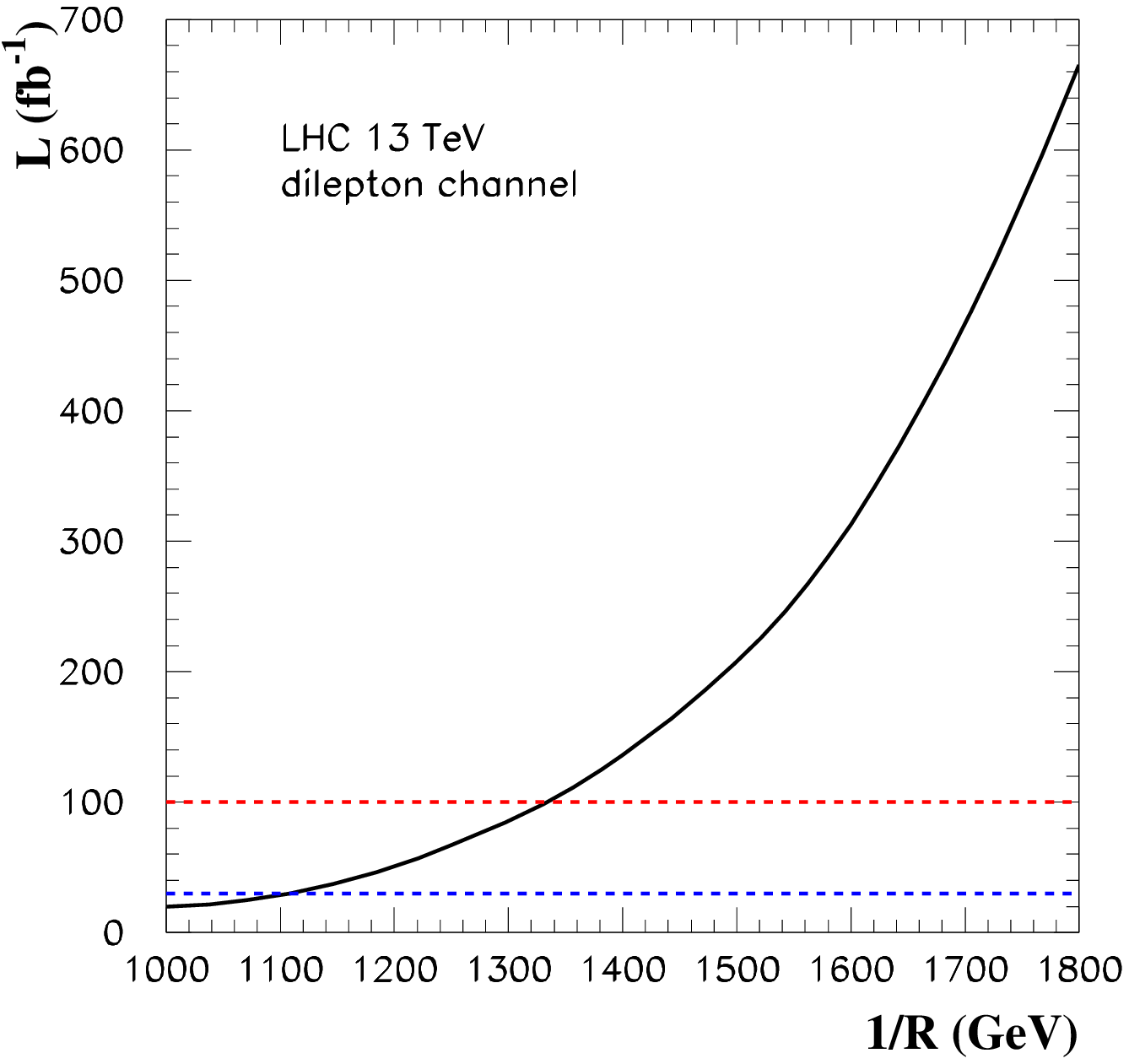}
  \caption{ Integrated luminosity required for a $5\sigma$ discovery
    of the six-dimension standard model (solid black curve) as a
    function of $B^{(1,1)}_\mu$ mass (left panel) and $1/R$ (right
    panel). Just for reference we show the lines for 30 and 100
    fb$^{-1}$. }
\label{fig:lepdisc}
\end{figure}

\subsection{Search for $(1,1)$ KK quarks}

We can also look for the six-dimensional standard model through the
single production of $(1,1)$ KK quarks as in Eq.~(\ref{eq:kkq}).  As
mentioned earlier, the added advantage of this mode is that it is more
model-specific since it requires a specific spectrum tied to a
particular structure of higher dimensional operators, making this
single production channel possible. Furthermore, the fact that the $(1,1)$ KK
quark decays to the dilepton resonance ($W_\mu^{3(1,1)}$ or
$B_\mu^{(1,1)}$) plus a hard jet allows the reconstruction
two states of the 6DSM spectrum in one decay channel. \smallskip

The main standard model backgrounds for this process are the Drell-Yan
and $W^+W^-/ZZ$ productions accompanied by a jet as well as top quark
pair production.  \smallskip

In order to extract the signal of KK $(1,1)$ quarks from the
background we required two hard charged leptons ($e^+e^-$ or
$\mu^+\mu^-$) as in Eq.~(\ref{eq:cuts:1}), as well as the presence of
an energetic jet in the event satisfying
\begin{equation}
| \eta_j| < 5 \;\;\;\;\hbox{and}\;\;\;\; p_T^j > 200 \hbox{ GeV} \;.
\end{equation}
Since the lepton pair originates from the decay of an on-shell
$B^{(1,1)}_\mu$ or $W^{(1,1)}_\mu$ we required the dilepton
invariant mass to be large
\begin{equation}
M_{\ell\ell} > 1.3 \times \frac{1}{R} \;,
\end{equation}
where $R$ is the compactification scale being probed. Furthermore, we
added the signal for the production of $D^{(1,1)}_-$, $D^{(1,1)}_+$,
$U^{(1,1)}_-$, and $U^{(1,1)}_+$ by requiring that the invariant mass
of the dilepton pair and the most energetic jet satisfies
\begin{equation}
M_{D^{(1,1)}_- } - 150 <  M_{\ell\ell j} < M_{U^{(1,1)}_+ } + 150
\hbox{ GeV,}
\end{equation}
where the masses are the ones corresponding to the scale $1/R$.
\smallskip

Assuming that just the SM background is observed in the channel given
by Eq.~(\ref{eq:kkq}), we depict in the left panel of Figure
~\ref{fig:lepbounds13} the attainable limits on the $(1,1)$ KK
production cross section for integrated luminosities of 30 fb$^{-1}$
(blue dashed line) and 100 fb$^{-1}$ (red dashed line) as well as the
6DSM expected cross section (solid black line). As we can see, using
this channel the LHC
run at 13 TeV has the potential of ruling out compactification scales
($1/R$) 1.4 TeV and 1.7 TeV for these integrated luminosities
respectively.\smallskip

The right panel of Figure ~\ref{fig:lepbounds13} displays the
integrated luminosity needed to establish a $5\sigma$ discovery of a
$(1,1)$ KK quark as a function of $1/R$. It is
interesting to notice that this channel can establish the 6DSM for
compactification scales ($1/R$) 1140 and 1360 GeV for integrated
luminosities of 30 and 100 fb$^{-1}$ respectively.  These values of
$R$ correspond to (1,1) KK quark masses around 2 and 2.3
TeV. Moreover, the reach in this channel is slightly larger than the
one in the dilepton channel for the same integrated luminosity.
\smallskip

\begin{figure}[htb!]
  \centering
  \includegraphics[width=0.45\textwidth]{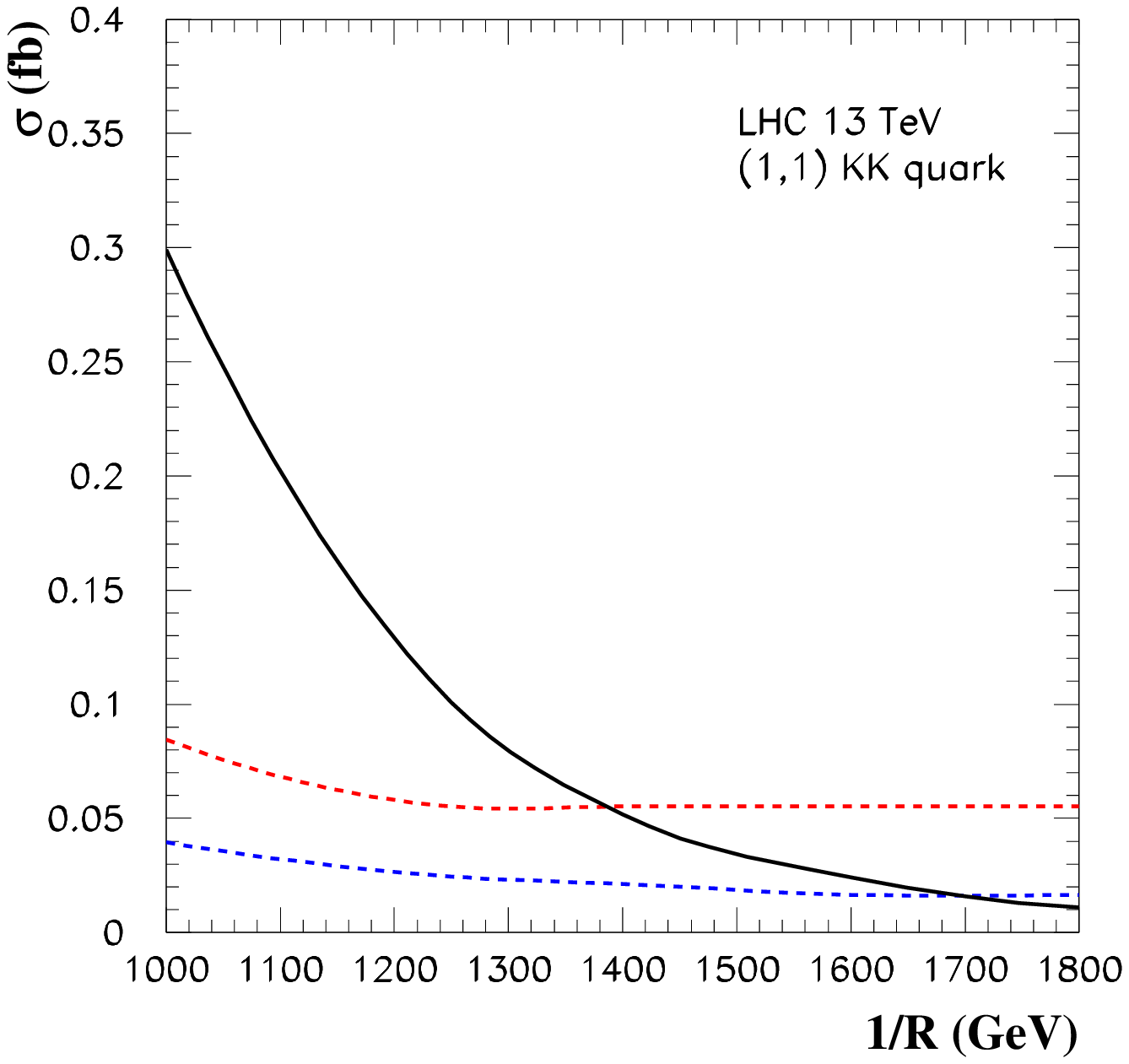}
  \includegraphics[width=0.45\textwidth]{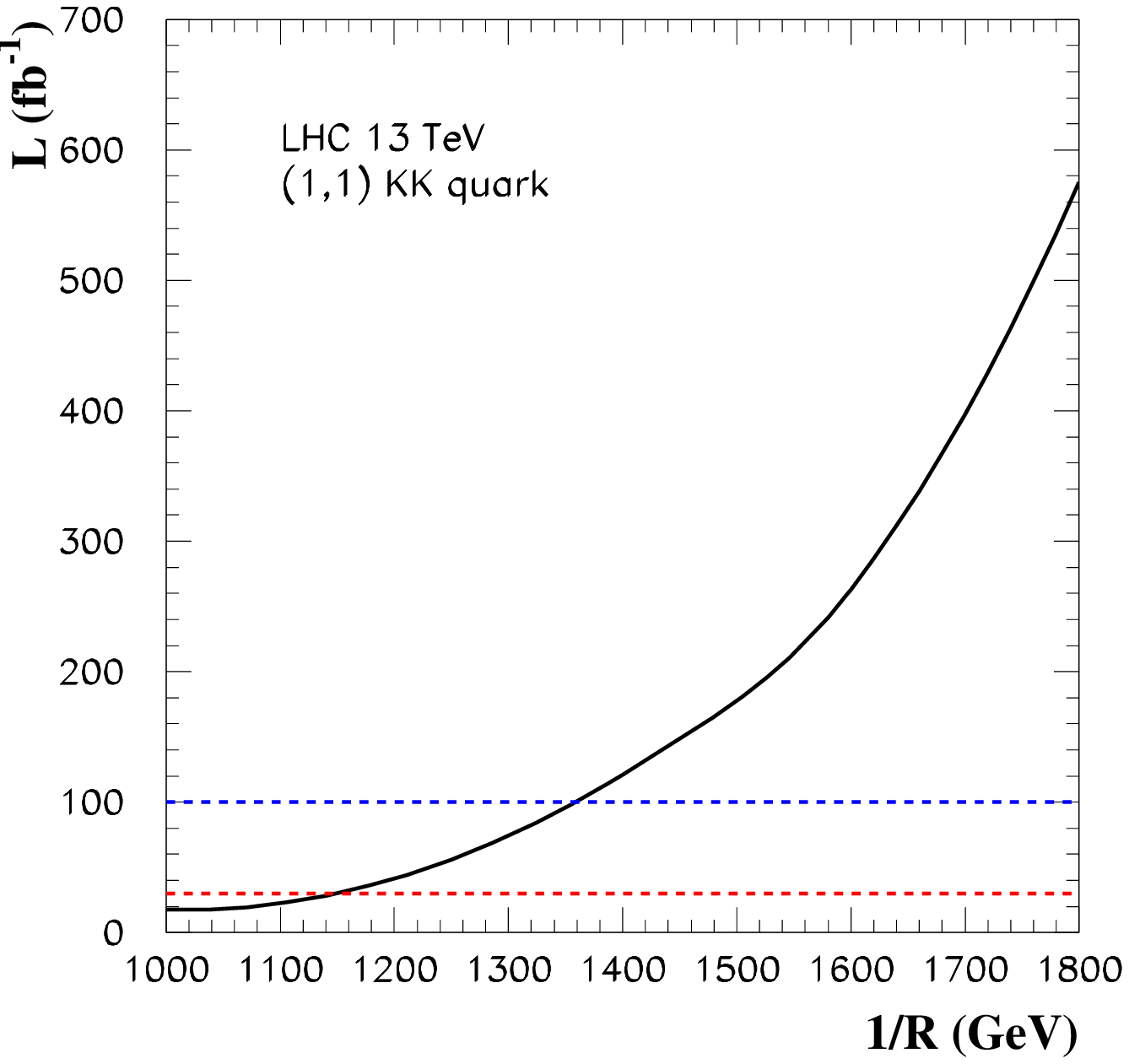}
  \caption{Left panel: 95 \% CL attainable limits on production cross
    section of $(1,1)$ KK quarks as function of $1/R$.  In this panel,
    the blue (red) dashed curve stands for the limit on the production
    cross section after cuts assuming an integrated luminosity of 30
    (100) fb$^{-1}$, while the solid black line stands for the
    expected production cross section within the six-dimensional
    standard model framework.  Right panel: Integrated luminosity
    required for a $5\sigma$ discovery of $(1,1)$ KK quarks (red
    curve) as a function of $1/R$. Just for reference we show the
    lines for 30 and 100 fb$^{-1}$. }
\label{fig:lepbounds13}
\end{figure}

\section{Conclusions}
\label{conclusions}

We have examined the present status and future potential of the LHC
bounds on two universal extra dimensions. The present limits
from the CMS available data we extracted in
Section~\ref{bounds} were obtained by simply using the 
  inclusive production of the $(1,1)$ excitations of the electroweak
gauge bosons, $W_\mu^{3(1,1)}$ and $B_\mu^{(1,1)}$ and their
subsequent decays to lepton pairs. The first excitations $(1,0)$ decay
mostly through KK-number conserving interactions resulting in low
$p_T$ tracks and missing $E_T$. The fact that the $(1,1)$ modes are
only $\sqrt{2}$ heavier than this so they must decay to SM states
through KK-number violating interactions, makes these modes easier to
search for at the LHC.  In fact, the bound extracted from the Run~I
data in these channels, $1/R>900~$GeV, is comparable to the indirect
bound obtained by using electroweak precision
measurements~\cite{Kakuda:2013kba}. \smallskip

We have also explored the LHC Run~II reach at $\sqrt{s}=13~$TeV both
in the dilepton 
resonance, as well as in the singly produced
$(1,1)$ quarks. Although the sensitivity in $1/R$ is similar in both
channels, the $(1,1)$ quark channel has the added advantage of being
more model specific when compared to the production of an 
vector resonance decaying to a pair of leptons. It is also interesting
that in this channel it would be possible to reconstruct not only the
dilepton resonance, but also the $(1,1)$ quark itself when the
dilepton is combined with the very hard jet. A more detailed
simulation of this reconstruction is needed which we leave for future
work.

\section*{Acknowledgments}
G.B and O.J.P.E. are supported in part by the Conselho Nacional de
Desenvolvimento Cient\'{\i}fico e Tecnol\'ogico (CNPq) and by
Funda\c{c}\~ao de Amparo \`a Pesquisa do Estado de S\~ao Paulo
(FAPESP). G.B. also acknowledges the hospitality of the Institute of
Advanced Study, Princeton during much of this work, and the generosity
of the Ambrose Monell Foundation during his stay there.  D.S. would
like to thank the instituto de F\'isica da USP for its kind
hospitality.


\end{document}